# Interplay of electron-lattice interactions and superconductivity in $Bi_2Sr_2CaCu_2O_{8+\delta}$

Jinho Lee[1], K. Fujita[1,2], K. McElroy[1,3], J.A. Slezak[1], M. Wang[1], Y. Aiura[4], H. Bando[4], M. Ishikado[2], T. Masui[5], J. -X. Zhu[6], A. V. Balatsky[6], H. Eisaki[4], S. Uchida[2], and J. C. Davis[1]

[1] *LASSP, Department of Physics, Cornell University, Ithaca, NY 14853, USA.*

[2] *Department of Physics, University of Tokyo, Tokyo, 113-8656, Japan.*

[3] *Materials Sciences Division, Berkeley National Lab., Berkeley, CA 94720, USA.*

[4] *AIST, 1-1-1 Central 2, Umezono, Tsukuba, Ibaraki, 305-8568, Japan.*

[5] *Department of Physics, Osaka, University, 1-1 Machikaneyama, Toyonaka, Osaka, 560-0043, Japan.*

[6] *T-Division, MS B 262, Los Alamos National Lab., Los Alamos, NM 87545, USA.*

**Formation of electron pairs is essential to superconductivity. For conventional superconductors, tunnelling spectroscopy has established that pairing is mediated by bosonic modes (phonons); a peak in the second derivative of tunnel current $d^2I/dV^2$ corresponds to each phonon mode[1-3]. For high-transition-temperature (high-$T_c$) superconductivity, however, no boson mediating electron pairing has been identified. One explanation could be that electron pair formation[4] and related electron-boson interactions are heterogeneous at the atomic scale and therefore challenging to characterize. However, with the latest advances in $d^2I/dV^2$ spectroscopy using scanning tunnelling microscopy, it has become possible**

to study bosonic modes directly at the atomic scale[5]. Here we report $d^2I/dV^2$ imaging[6-8] studies of the high-$T_c$ superconductor Bi$_2$Sr$_2$CaCu$_2$O$_{8+\delta}$. We find intense disorder of electron-boson interaction energies at the nanometre scale, along with the expected modulations in $d^2I/dV^2$ (refs 9,10). Changing the density of holes has minimal effects on both the average mode energies and the modulations, indicating that the bosonic modes are unrelated to electronic or magnetic structure. Instead, the modes appear to be local lattice vibrations, as substitution of $^{18}$O for $^{16}$O throughout the material reduces the average mode energy by approximately 6 per cent - the expected effect of this isotope substitution on lattice vibration frequencies[5]. Significantly, the mode energies are always spatially anticorrelated with the superconducting pairing-gap energies, suggesting an interplay between these lattice vibration modes and the superconductivity.

Strong-coupling superconductivity theory[1-3,11], allowd the impact of electron-phonon interactions on the superconducting density of states, DOS($E$), at energy $E = \Delta + \Omega$ to be predicted ($\Delta$ is the superconducting energy gap and $\Omega$ the phonon energy). McMillan and Rowell used $d^2I/dV^2$ measurements on planar normal-insulator-superconductor tunnel junctions[1] to reveal these signatures at boson energies $\Omega = E - \Delta$, identifying them with independently determined phonons of energy $\Omega$. The electron-phonon spectral function measured from these $d^2I/dV^2$ spectra yielded the correct superconducting transition temperature - a milestone of twentieth-century physics.

By contrast, no consensus exists on the electron pairing mechanism of high-$T_c$ superconductivity. One reason is that the extensive studies of bosonic modes[12-17] and related electronic self-energy changes[18-26] have not resulted in the unambiguous identification of a boson mediating the pairing. Effects of electron-boson interactions (EBI) on electronic self-energies are most widely studied[18-26] via angle resolved photoemission (ARPES). In the $\Delta=0$ or nodal direction of momentum space $\vec{k}//(\pi/a_0,\pi/a_0)$, where $a_0$ is the unit cell dimension, sudden changes or "kinks" in quasiparticle dispersion $E(k)$ occur between 50meV and 80meV below the Fermi energy[18-21]. Several studies have addressed the issue of whether these "kinks" might be due to magnetic interactions[19-20]. On the other hand, it has been proposed that, because of the doping independence of their energies, the "kinks" are due to electron-phonon interactions[21]. In the antinodal directions $\vec{k} \approx (\pi/a_0,0)$ where $\Delta$ is maximal, self energy changes[22-25] also occur between 50meV and 90meV below the Fermi energy. But here the effects are thought to occur at $E=\Delta+\Omega$. These modes have been discussed in terms of both magnetic interactions[22-24] and electron-phonon[25] interactions. Effects of $^{16}O/^{18}O$ isotope substitution have been found primarily on states near $\vec{k} \approx (\pi/a_0,0)$, providing evidence for electron interactions with lattice modes[26] (to be discussed in more detail below). Another EBI study technique is superconductor-insulator-superconductor (SIS) conductance measurements in break junctions[27]. Despite the challenge of interpreting SIS spectra in terms of the absolute value of $\Omega$ (they are not a direct measure of the superconducting DOS[28]), clear EBI features are detected; they are analyzed in terms of magnetic modes. A final probe of copper oxide EBI is optical spectroscopy[29,30]; it reveals self-energy changes that have been ascribed to magnetic interactions via a sharp mode[29] or a broad continuum[30]. Evidently, a definite conclusion for the identity of any pairing related EBI has proven elusive.

One reason for this situation might be that some key element of the EBI phenomenology has, so far, gone undetected. For example, if electron pairing[4] and any related EBI were disordered at the atomic scale, the techniques described above could not yield a complete description of the relevant nanoscale EBI phenomena because they average over space. For such reasons, atomic-resolution $d^2I/dV^2$ spectroscopy of copper oxides has recently become the focus of considerable theoretical interest[6-10]. Three of the proposed applications are: (1) if the position(r) dependence of energy-gap disorder $\Delta(\vec{r})$ were due to atomic scale pair-potential disorder[4], pairing-related EBI could be examined directly at that scale, (2) studies of specific local bosonic modes (unrelated to pairing) at defects or impurity/dopant atoms[6-8] could improve our understanding of EBI in copper oxides and, (3) Fourier transform $d^2I/dV^2$ imaging (dubbed FT-IETS)[10] may help distinguish between different bosons involved in EBI. However, until now, severe technical constraints (see Supplementary Information) have prevented implementation of $d^2I/dV^2$ imaging of the necessary spatial/energetic precision.

Here we report atomic-resolution $d^2I/dV^2$ – imaging studies of a high-$T_c$ superconductor $Bi_2Sr_2CaCu_2O_{8+\delta}$ (Bi-2212). We use floating-zone grown single crystals cleaved between the BiO planes in cryogenic ultra-high vacuum and immediately inserted into the STM head at 4.2 K. Figure 1a shows a topographic image of a typical

BiO surface, while the inset shows the typical local density of states (*LDOS*) measured via a differential conductance $dI/dV$ ($E = eV$) spectrum. Here we focus on the nanoscale spatial/energetic properties of the ubiquitous features at $|E|>\Delta$ in the *LDOS* (for example, see arrows in Fig. 1a inset).

Figure 1b gives typical examples of **$dI/dV(E)$** spectra measured at different locations of Fig. 1a. The vast majority of these spectra exhibit peaks in $d^2I/dV^2(E)$ occurring at the point of maximum slope in $dI/dV$ for $|E|>\Delta$ (arrows in Fig. 1b). Examples of the directly measured magnitude of the features in **$d^2I/dV^2$** are shown in Fig. 1b, where the horizontal axis has first been converted to $\omega = E - \Delta$ (where $\Delta$ is the distinct local gap magnitude for each spectrum). We provisionally consider these features as possible strong-coupling superconductivity[1-3] signatures of EBI (see Supplementary Information).

The $dI/dV(\vec{r},E)$ and $d^2I/dV^2(\vec{r},E)$ are simultaneously imaged with atomic resolution and register. From the former, the gap-map $\Delta(\vec{r})$ is derived (Fig. 1c). From the latter, the energies $\Pi(\vec{r})$ at which peaks in **$d^2I/dV^2(\vec{r},E)$** occur are measured. Within the context of strong-coupling superconductivity theory[1-3,9], the local boson interaction-energies would then be given by $\Omega(\vec{r}) = \Pi(\vec{r}) - \Delta(\vec{r})$. Some systematic uncertainties may exist in the precision with which this process yields the absolute mode energy (especially for complex band structures). Nevertheless, the mean mode energy is estimated from the $\Omega(\vec{r})$ to be $\overline{\Omega} \approx 52$ meV with a statistical spread of $\pm 8$ meV. This is within the range of mode energies for antinodal EBI reported in photoemission studies[22-25]. However, it is from analysis of the $\Omega(\vec{r})$ images (Fig. 1d, Fig. 3d-f) that a very different perspective on the EBI of Bi-2212 emerges. We see immediately that boson energies $\Omega(\vec{r})$ are heterogeneous at the ~2nm scale with $40 meV < \Omega(\vec{r}) < 65 meV$, spanning the range of antinodal mode energies from photoemission[22-25]. The necessity of atomic-resolution $d^2I/dV^2(\vec{r}, E)$ – imaging studies to fully explore EBI signatures in high-$T_c$ superconductors becomes manifest here.

In theory[9,10], spatial modulations of $d^2I/dV^2(r, E)$ near $E = \Omega + \Delta$ can contain key information about the EBI. But we detect no spatially periodic structure in unprocessed $d^2I/dV^2(E)$ images in this energy range. This can be understood, however, because $\Delta(\vec{r})$ is so strongly disordered (Fig. 1c) that any EBI[10] effects in $d^2I/dV^2(\vec{r}, \Omega + \Delta(\vec{r}))$ would be spatially scrambled. To search for these effects, each $d^2I/dV^2(E)$ must therefore be shifted to its bosonic energy scale $\omega(\vec{r}) = E - \Delta(\vec{r})$, a process we refer to as 'gap referencing'. This converts the unprocessed $d^2I/dV^2(\vec{r}, E); |E| > \Delta$ data into a new series of $d^2I/dV^2(\vec{r}, \omega)$ images. These vary little (except in intensity) within the energy range 40meV<$\omega$<65meV where peaks in $d^2I/dV^2(\omega)$ are detectable (See Supplementary. Fig. 1). Remarkably, they all exhibit the same spatial modulations, which did not exist in unprocessed $d^2I/dV^2(\vec{r}, E)$ data before gap referencing (see Supplementary Fig. 1).

To enhance their spatial contrast, we sum the $d^2I/dV^2(\vec{r},\omega)$ images over the boson energy range where EBI are detected: $\Gamma(\vec{r}) = \sum_{\omega=40meV}^{\omega=65meV} d^2I/dV^2(\vec{r},\omega)$. This energy average produces lower spatial noise compared to a single energy map. A typical resulting $\Gamma(\vec{r})$ is shown in Fig. 2a; the modulations are parallel to the Cu-O bond directions and have wavelengths ~5$a_0$ with correlation length of ~50Å. We then use $\Gamma(\vec{q})$, the Fourier transform of $\Gamma(\vec{r})$, in Fig. 2b to determine that the modulation wavevectors are $\vec{p}_1 \approx 2\pi/a_0[(0.2,0);(0,0.2)] \pm 15\%$. Within the theoretical models[9,10] such $d^2I/dV^2(\vec{r},\omega)$ modulations are created when electronic states, renormalized by EBI, are scattered by disorder. When scattering is modelled as due to atomic-scale variations in the pair potential[4], the predictions for $d^2I/dV^2(\vec{r},\omega)$ modulations[10] are qualitatively consistent with data in Fig. 2, if the boson is a lattice vibration mode.

Next we study these $d^2I/dV^2$ signatures of EBI at a sequence of different hole densities per CuO$_2$, $p$: $p \sim 0.12 \rightarrow p \sim 0.24$. In Fig. 3a-c we show that the average superconducting energy gap $\overline{\Delta}$ decreases from $\sim 60meV \rightarrow \sim 20meV$ with increasing doping, as expected. In strong contrast, Fig. 3d-f shows that, although changes occur in spatial correlations of $\Omega(\vec{r})$, no change is detectable in the average boson energy $\overline{\Omega}$. In fact, histograms of $\Delta$ and $\Omega$ measured on five samples at different dopings (Fig. 4) reveal that, while the distributions of $\Delta$ evolve rapidly with doping, those of $\Omega$ appear unchanged: $\overline{\Omega} = 52 \pm 1meV$ for all dopings. We emphasize that the doping independence of $\overline{\Omega}$ is not because the average energy $\overline{\Pi}$ of the $d^2I/dV^2$ peak is unchanged with doping; $\overline{\Pi}$ changes from $\overline{\Pi} \cong \pm 115$meV to $\overline{\Pi} \cong \pm 70$meV over the doping range. It is the *difference,* $\overline{\Omega}$**,** between $\overline{\Pi}$ and $\overline{\Delta}$ which remains constant. Furthermore, we find the $d^2I/dV^2$ – modulation wavevectors $\vec{p}_1$ are the same for all gap-referenced $\Gamma(\vec{r})$ at all dopings (see Fig. 2c).

Thus, both $d^2I/dV^2(r,\omega)$ modulations and $\overline{\Omega}$ are independent of doped hole-density. Which boson could exhibit such highly doping-independent EBI characteristics? The 'resonant' spin-1 magnetic excitation mode[15] appears inconsistent because its energy is 43 meV in Bi-2212 but, more importantly, is believed to be strongly doping dependent. The incommensurate, dispersive, spin density wave modes[16,17] also appear inconsistent because of their characteristic strong energy- or doping-dependences. By contrast, because energies of lattice-vibration modes change little with doping, they are logical candidates for the boson detected by $d^2I/dV^2$ imaging.

Indeed, electron-lattice interactions are well known to influence copper oxide superconductivity. For example, the energy of a phonon mode at momentum transfer $\vec{Q} \approx 2\pi/a_0[(0.25,0);(0,0.25)]$ diminishes rapidly towards 50meV upon cooling into the superconducting state[12-14], as might be expected for strong phonon interactions with superconducting quasiparticles. Furthermore, studies have revealed an unusual $^{16}O/^{18}O$ isotope substitution effect[26] on the electronic structure; the data point to maximum influence of lattice modes on the high energy (*E*~-250meV) electronic structure near the antinodes, with a much weaker impact above $T_c$ and at low energy. Although these results are not consistent with the simplest Eliashberg picture of electron phonon interactions[26], they do represent evidence for interactions between states near $\vec{k} \approx (\pi/a_0, 0)$ and lattice vibrational modes. Furthermore, there have been detailed photoemission studies [25] of antinodal quasiparticles coupling to a bosonic mode - ascribed to a $B_{1g}$ Cu-O bond-buckling phonon from theoretical analysis. Taken in combination, these studies point to interactions between antinodal quasiparticles and Cu-O related lattice vibrations as influencing high temperature superconductivity - although the precise implications for electron pairing mechanism remain uncertain.

To test the hypothesis that bosons detectable by $d^2I/dV^2$ – imaging techniques in $Bi_2Sr_2CaCu_2O_{8+\delta}$ are lattice vibration modes, we prepared crystals in which the normal $^{16}O$ was completely substituted by $^{18}O$ (as verified by frequency shifts detected in Raman spectroscopy). Figure 5a provides the comparisons between the distributions of $\Omega(\vec{r})$ and $\Delta(\vec{r})$ in different samples containing complete substitutions of the two oxygen isotopes. For each sample, we take $\Delta(\vec{r})$ and $\Omega(\vec{r})$ and construct a two-dimensional histogram of the frequency of occurrence of spectra with a given pair of values $(\Delta, \Omega)$. Each histogram is peaked along the vertical axis at the most common gap energy $\overline{\Delta}$ and along the horizontal axis at the most common boson energy $\overline{\Omega}$. Comparison between $^{16}O$ $\Delta\Omega$-histogram (blue) and the $^{18}O$ $\Delta\Omega$-histogram (red) reveals immediately that $\overline{\Omega}$ for $^{18}O$ shifts downwards by several meV compared to that of $^{16}O$. A quantitative analysis in Figure 5b shows the distribution of boson energy $\Omega(r)$ in two different samples containing complete substitutions of the two oxygen isotopes: $^{16}O$ in blue and $^{18}O$ in red. We find that the shift of $\overline{\Omega}$ upon substitution of $^{16}O$ by $^{18}O$ is -3.7±0.8 meV. These results are found equally true for both filled $E=-(\Delta+\Omega)$ and empty $E=+(\Delta+\Omega)$ states, as expected for EBI in strong-coupling superconductivity theory[2,3]. Consequently, substitution of $^{18}O$ for $^{16}O$ reduces the mean boson energy scale $\overline{\Omega}$ of EBI by $6\% (\approx 1-\sqrt{16/18})$ - as expected for lattice vibrational modes involving the O atom.

These $d^2I/dV^2$ – imaging studies alter several existing concepts of the EBI problem in $Bi_2Sr_2CaCu_2O_{8+\delta}$. We demonstrate that it is modes involving lattice vibrations which generate the $d^2I/dV^2$ features in tunnelling. Further, the diminishing intensities of the $d^2I/dV^2$ peaks with coherence peak height (Fig. 1b) along with the necessity for gap referencing (Fig. 2), signify the primary involvement of the antinodal states ($E \sim \Delta, \vec{k} \sim (\pi,0)$) in the interactions. The mean mode energy is $\overline{\Omega} \approx 52$ meV with a statistical spread of ±8meV. These modes exhibit intense atomic scale disorder of interaction energies $\Omega(\vec{r})$ whose doping independence (Fig. 4b) points to a population native to the crystal. And perhaps most significantly, the $\Omega(\vec{r})$ are spatially anticorrelated with the superconducting energy gap disorder $\Delta(\vec{r})$ at all dopings (Fig. 3) and for both oxygen isotopes (Fig. 5a). Finally, in Fig. 5c we show the normalized correlations between the dopant atom locations $O(\vec{r})$ and both $\Omega(\vec{r})$ and $\Delta(\vec{r})$. Whereas the zero-displacement correlations $O(\vec{r}) : \Delta(\vec{r}) \approx +0.35$ are as expected, and the $\Omega(\vec{r}) : \Delta(\vec{r}) \approx -0.30$ correlations are consistent with Fig. 3, we find that $\Omega(\vec{r})$ and $O(\vec{r})$ are uncorrelated. Therefore, correlations between $\Omega(\vec{r})$ and $\Delta(\vec{r})$ cannot be occurring trivially, through a similar effect of dopant disorder on both. A direct atomic-scale influence of $\Omega(\vec{r})$ on $\Delta(\vec{r})$ (or vice versa) is implied.

Taken together, these data present some intriguing new possibilities. The first is that superconducting energy gap disorder $\Delta(\vec{r})$ is a consequence of heterogeneity in the pairing potential caused by disorder in the frequencies and coupling constants of pairing-related vibrational modes[31]. But the strong dependence of superconducting electronic structure on hole density while the $\Omega(\vec{r})$ distributions remain unchanged (Fig. 4) appears to argue against this point of view. A second possibility is that the $d^2I/dV^2$ features are unconnected to pairing-related EBI - perhaps occurring because of inelastic stimulation of vibrational modes within the tunnel barrier itself[32] or because of non-pairing-related electron lattice interactions. The primary difficulty here is that the ubiquitous anticorrelation between $\Omega(\vec{r})$ and $\Delta(\vec{r})$ cannot be explained trivially within such scenarios. A third possibility is that the $d^2I/dV^2$ features represent electron-lattice interactions related to a competing electronic ordered state (see, for example, Ref. 13), and that the anticorrelation between $\Omega(\vec{r})$ and $\Delta(\vec{r})$ occurs because of this competition. To help to distinguish between these possibilities, an atomic-scale version of the McMillan-Rowell procedure[1] may now become necessary.

**Figure 1 Atomic-resolution $d^2I/dV^2$ $(\vec{r},E)$ imaging of electron-boson interactions in Bi$_2$Sr$_2$CaCu$_2$O$_{8+\delta}$.**

**a,** Typical topographic image of one of the surfaces under study(Z, surface height). The inset shows the characteristic $dI/dV$ spectrum which is proportional to the *DOS*. The ubiquitous features occurring at $E > \Delta$ (where $\Delta$ is the superconducting energy gap), whose spatial and energetic structure are of central interest in this paper, are indicated by arrows. **b,** Examples of $dI/dV$ spectra in different regions of surface shown in **a**; left, the peaks in $d^2I/dV^2$ occur at the points of maximum slope of $dI/dV$ for $E > \Delta$ as indicated by arrows; right, examples of directly measured peaks in $d^2I/dV^2$ from the identical locations as the same-coloured $dI/dV$ spectra in **b**. **c,** The image of superconducting energy gap values, or gap-map $\Delta(\vec{r})$, on surface in **a**. **d,** image of distribution of boson energies $\Omega(\vec{r}) = \Pi(\vec{r}) - \Delta(\vec{r})$ (where $\Pi(\vec{r})$ are bias energies at which $d^2I/dV^2$ peaks occur) on surface in **a**. This analysis scheme to find $\Omega(\vec{r})$ has proven reliable and repeatable on numerous Bi-2212 samples at a wide range of dopings and with different oxygen isotopes. In all cases, it gives statistically indistinguishable results for both the filled and empty states: $E < E_F$ and $E > E_F$.

**Figure 2 Ubiquitous quasi-periodic spatial modulations in $d^2I/dV^2(\vec{r},\omega)$ signals, after gap referencing.**

a, We define the local bosonic energy scale as $\omega(\vec{r}) = E - \Delta(\vec{r})$. A typical map of $d^2I/dV^2(\vec{r},\omega)$ modulations integrated over energy $\Gamma(\vec{r}) = \sum_{\omega=40meV}^{\omega=65meV} d^2I/dV^2(\vec{r},\omega)$ then reveals directly that the quasi-periodic spatial $d^2I/dV^2(\vec{r},\omega)$ modulations are parallel to the Cu-O bond directions, have wavelengths of $5a_0$ and correlation length of ~ 50 Å. These phenomena occur in all gap-referenced $\Gamma(\vec{r})$.

**b,** The characteristic wavevectors of these $d^2I/dV^2(\vec{r},\omega)$ modulations, as indicated directly by the arrow in $\Gamma(\vec{q})$, the Fourier transform (FT) of $\Gamma(\vec{r})$, are $\vec{p}_1 \approx 2\pi/a_0[(0.2,0);(0,0.2)]\pm 15\%$. They are dispersionless within our resolution.

**c,** The doping dependence of these $\vec{p}_1$ are analyzed by plotting the magnitude of $\Gamma(\vec{q})$ along the line (0,0) to (0,2π) as shown in **b**. We find that very similar modulation wavevectors (black arrows) occur at all dopings (FFT, fast Fourier transform; a.u., arbitrary units).

**Figure 3 Doping dependence of electron-boson interactions of $Bi_2Sr_2CaCu_2O_{8+\delta}$.**

**a-c,** Gapmap $\Delta(\vec{r})$ for three values of $p$, respectively ~ 0.12, ~ 0.18 and ~ 0.24. **d-f,** Simultaneously determined $\Omega(\vec{r})$ images. Note the colour bar is reversed here to show directly how higher $\Omega$ is correlated to lower $\Delta$ and vice versa. Also, we can see that spatial correlations of $\Delta(\vec{r})$ and $\Omega(\vec{r})$ both change together with doping. Regions of **a** and **d** which are black are where neither the value of $\Delta$ nor $\Omega$ can be determined because there are no coherence peaks or $d^2I/dV^2$ – peak features (see blue spectra in Fig. 1b).

**Figure 4 Doping dependence of energy gap histograms and boson energy histograms in $Bi_2Sr_2CaCu_2O_{8+\delta}$.**

**a,** Histograms of measured energy gaps $\Delta$ from a sequence of samples with different dopings, black being strongly overdoped and blue strongly underdoped. We see clearly that $\overline{\Delta}$ falls rapidly with rising doping and distribution of $\Delta$ broadens and flattens: there can be little doubt that the doping is indeed changing. **b,** Histograms of measured boson energies $\Omega$, from $d^2I/dV^2$ – imaging measurements performed simultaneously with **a**. Within the uncertainly, neither the distribution nor mean value of $\overline{\Omega} = 52 \pm 1$ meV are influenced by doping. Furthermore, since the same value of $\Omega(\vec{r})$ is associated with different absolute values of $\Delta(\vec{r})$ at different dopings; the most plausible explanation is that the doping-independent distributions of $\Omega$ are inherent to the crystal. We note that in our $> 10^6$ atomically resolved $dI/dV$ and $d^2I/dV^2$ spectra in this study, the minimum in $dI/dV$ always occurs near $\Omega_{dip} = \pm 26$ meV at all dopings (See Suppl. Fig. 2).

**Figure 5** $^{18}$O/$^{16}$O **isotope effects on** $d^2I/dV^2(\vec{r},\omega)$ **spectra and the distribution of boson energies.**

The substitution of $^{16}$O by $^{18}$O was demonstrated via Raman spectroscopy of both in-plane and out-of-plane oxygen vibrational mode frequencies. **a,** Two dimensional $\Delta\Omega$-histograms of the frequency of occurrence of a given pair of $\Delta,\Omega$ values in a single spectrum. Data for $^{16}$O is blue and $^{18}$O is red. Although the $\Omega$ vary in a fashion correlated with $\Delta$, the shift in $\Omega$ with substitution of $^{16}$O by $^{18}$O is downwards by several meV. The vertical shift of 5.6 meV in $\overline{\Delta}$ between samples occurred inadvertently because the hole density, as determined independently from $T_c$ ($^{16}T_c$~76K, $^{18}T_c$~88K), was slightly different in the two samples. As far as we know at present, no importance should be attributed to this shift. **b,** The histograms for all values of $\Omega$ for samples with $^{16}$O (blue) and $^{18}$O (red). The average shift of energy with isotope substitution is -3.7±0.8 meV. We find this same shift if $\Omega$ distributions are measured for both filled ($E<E_F$) and empty ($E>E_F$) states. **c,** The normalized correlations between the dopant atom locations $O(\vec{r})$ and both $\Omega(\vec{r})$ and $\Delta(\vec{r})$. While zero-displacement dopant-gap-map $O(\vec{r}) : \Delta(\vec{r}) \approx +0.35$ as expected, and the $\Omega(\vec{r}) : \Delta(\vec{r}) \approx -0.30$ consistent with Fig. 3, we find that $\Omega(\vec{r})$ and $O(\vec{r})$ are uncorrelated (solid black line).

We acknowledge and thank Ar. Abanov, P.W. Anderson, N. Ashcroft, T. P. Devereaux, T. Egami, M. Eshrig, C. Henley, P. J. Hirschfeld, A. Lanzara, D. -H. Lee, P. Littlewood, D. Morr, K. A. Müller, J. Rowell, M. R. Norman, J. Orenstein, T. M. Rice, D. J. Scalapino, Z. -X. Shen, C. M. Varma & A. Zettl for helpful discussions and communications. This work was supported by an LDRD from Los Alamos National Laboratory, by Cornell University and by the ONR.

Correspondence and requests for materials should be addressed to J. C. D. (email: jcdavis@ccmr.cornell.edu)

**Supplementary Materials**

**Ultra low vibration laboratory**

To spatially image $d^2I/dV^2(\vec{r}, E = eV)$ with equal signal-to-noise ratio as that in high-resolution $dI/dV(\vec{r}, E)$ imaging is technically challenging. Consider that the pioneering measurements of Stipe et al[5] could take up to 10 hours per individual spectrum to detect inelastic peaks in $d^2I/dV^2(E)$; at such a rate $256^2$ pixel spatial image of $d^2I/dV^2(\vec{r}, E)$, would take ~ 100 years to acquire. To reduce acquisition time sufficiently for atomic resolution $d^2I/dV^2(E)$-imaging to become practical for copper-oxide studies, new ultra low vibration laboratories were constructed. They deeply suppress vibration levels at the STM head allowing each high-resolution $d^2I/dV^2(E)$ spectrum (e.g. Fig. 1b) to be measured every ~ 10 seconds. Complete $d^2I/dV^2(\vec{r}, E)$ data sets can now be acquired in about a month and, for this study, > $10^6$ atomically resolved and registered $d^2I/dV^2(\vec{r}, E)$ were acquired over a three year period.

**Measuring nanoscale EBI by $d^2I/dV^2$ – imaging**

Our $d^2I/dV^2$ - imaging studies of Bi-2212 may be interpreted as atomic-resolution imaging of EBI for the following reasons: (i) the intensity in $d^2I/dV^2$ – peaks depends ~linearly on the height of the superconducting coherence peak at $E = \Delta$ (see Fig. 1b); this is as expected for the strong-coupling superconductivity $d^2I/dV^2$ - signature of EBI since the strength of this feature is derived from the singularity in the *LDOS* at the gap energy[2,3], (ii) the average $d^2I/dV^2$ – peak energy $\overline{\Pi}$ **c**hanges from $\overline{\Pi} = -115$ meV to $\overline{\Pi} = -70$ meV with increasing doping while it is the difference between $\overline{\Pi}$ and $\overline{\Delta}$ which remains constant at $\overline{\Omega}$; the implication being that there is indeed a bosonic mode of fixed mean energy $\overline{\Omega}$ which is causing self-energy changes at $\overline{\Pi} = \overline{\Delta} + \overline{\Omega}$, with only $\overline{\Delta}$ varying with doping, (iii) no modulations are detected in the unprocessed $d^2I/dV^2(\vec{r},E): E > \Delta$ data before gap-referencing whereas clear spatial modulations are detected in all $d^2I/dV^2(\vec{r},\omega)$ after gap referencing (Suppl. Fig. 1); the implication being that predicetd[10] $d^2I/dV^2(\vec{r})$ modulations occur throughout – but remain undetected due to gap disorder and, (iv) $^{18}O/^{16}O$ isotope substitution shifts the whole $d^2I/dV^2(\vec{r},\omega)$ structure including $\overline{\Omega}$ values down in energy by ~ 6%, as expected if the bosonic mode is a lattice vibration involving the O atom.

**Supplementary Figure 1 Emergence of the spatial modulations in the gap-referenced $d^2I/dV^2(\vec{r},\omega)$ near $\omega = \Omega_0$.**

**a,** Unprocessed images of $d^2I/dV^2(r,E)$ before gap-referencing near $E \sim \Pi = \Delta + \Omega_0$, (near energy where $d^2I/dV^2$ is maximum outside the gap) and their 2D Fourier transforms. No clear modulation is visible in the real space image, nor in the 2D Fourier transform images. **b,** Images of $d^2I/dV^2(\vec{r},\omega)$ and their 2D Fourier transform images near $\omega \sim \Omega_0$, after gap-referencing. Clear modulations exist in both real space images and 2D Fourier transform images.

**Supplementary Figure 2 Doping dependence of gap histograms and dip histograms in $Bi_2Sr_2CaCu_2O_{8+\delta}$.**

**a,** Histograms of $\Delta$ from samples with different dopings. The same as in Fig. 4a.
**b,** Histograms of $E_{dip}$ where $dI/dV(r,E)$ minimum occurs outside the gap. These histograms are from the identical $d^2I/dV^2(r,E)$-imaging studies as in Fig. 4. Mean values of $E_{dip}$ occur near $\Omega = \pm 26$ meV for $^{16}O$ samples at all dopings.

**Supplementary Figure 3 Comparison of crystal quality between $^{16}O$ and $^{18}O$.**
We show typical topographic images taken during the studies reported in the text. One is of crystal with $^{16}O$ and the other with $^{18}O$. We have observed no diminution whatsoever in crystal quality due to the $^{18}O$ substitution process.

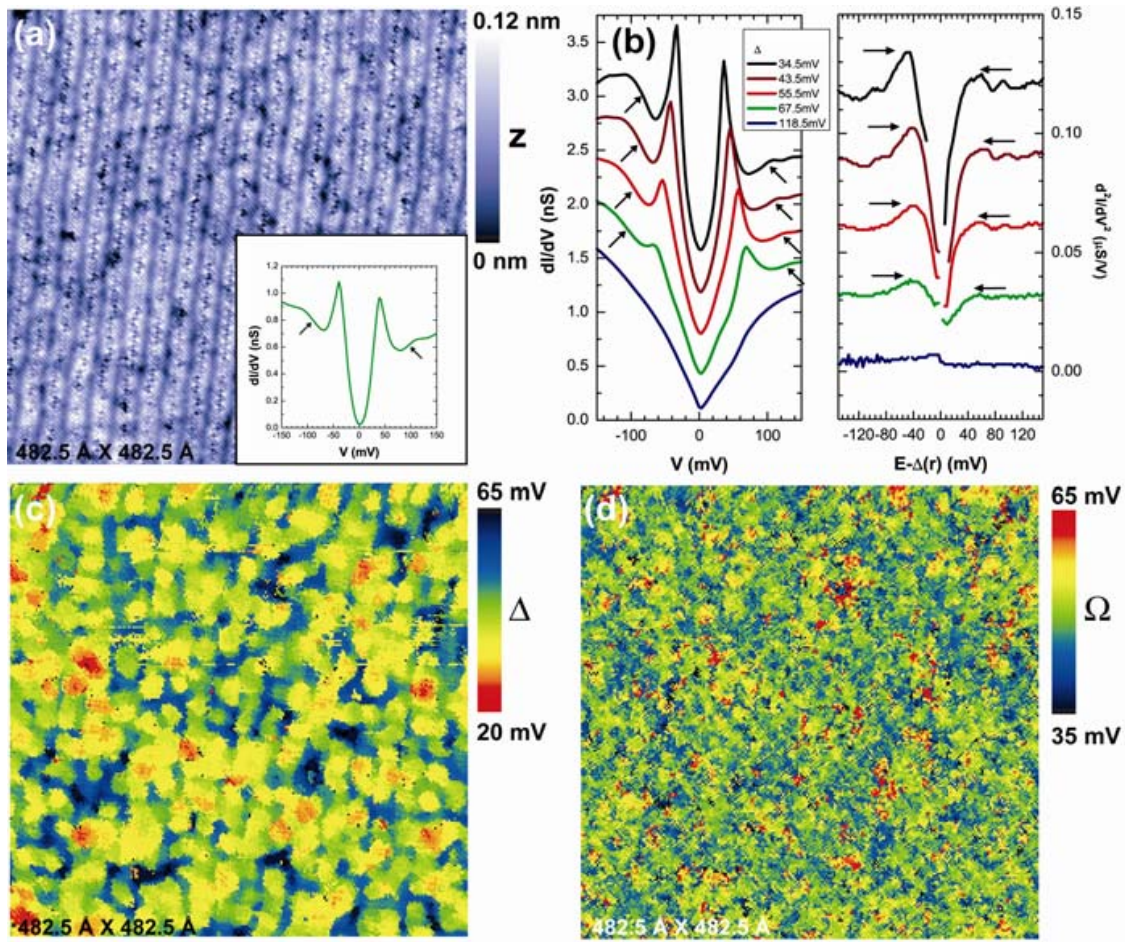

**Figure 1.**

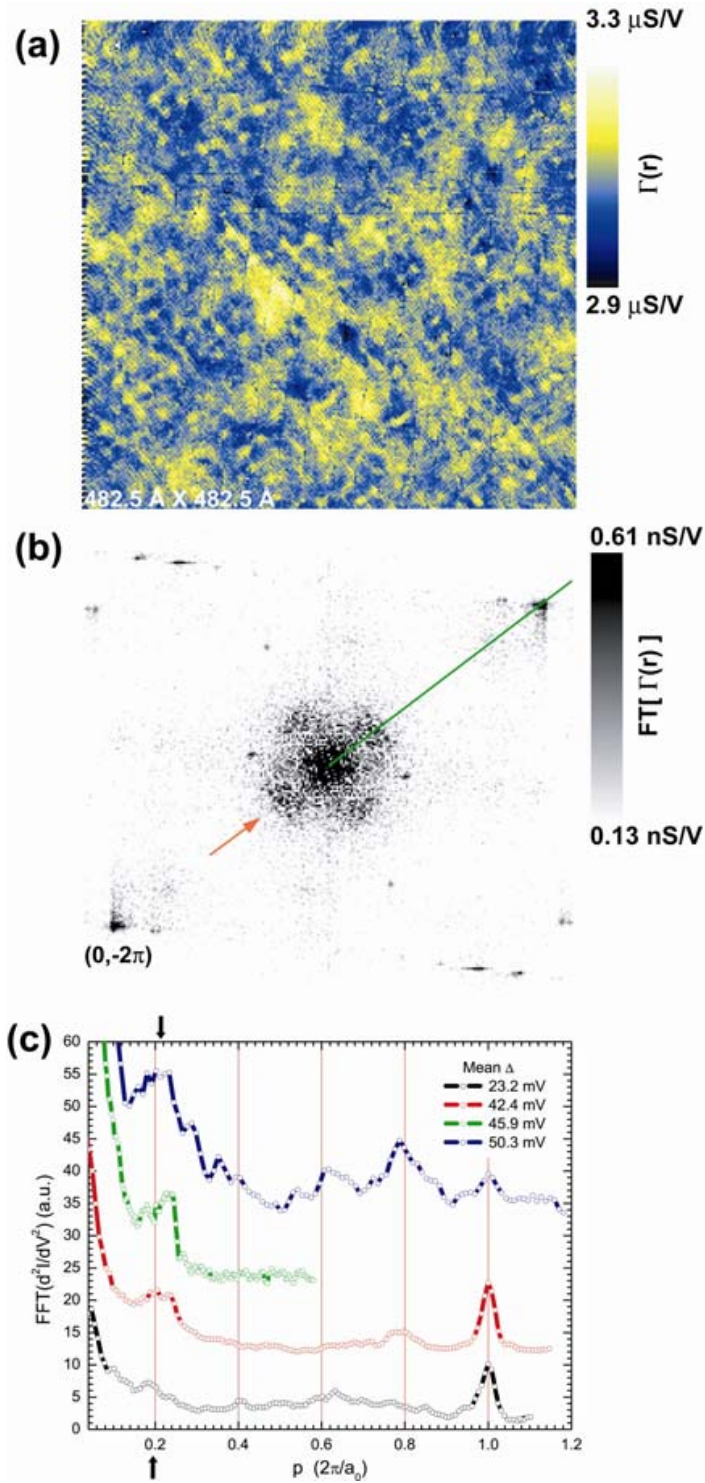

**Figure 2.**

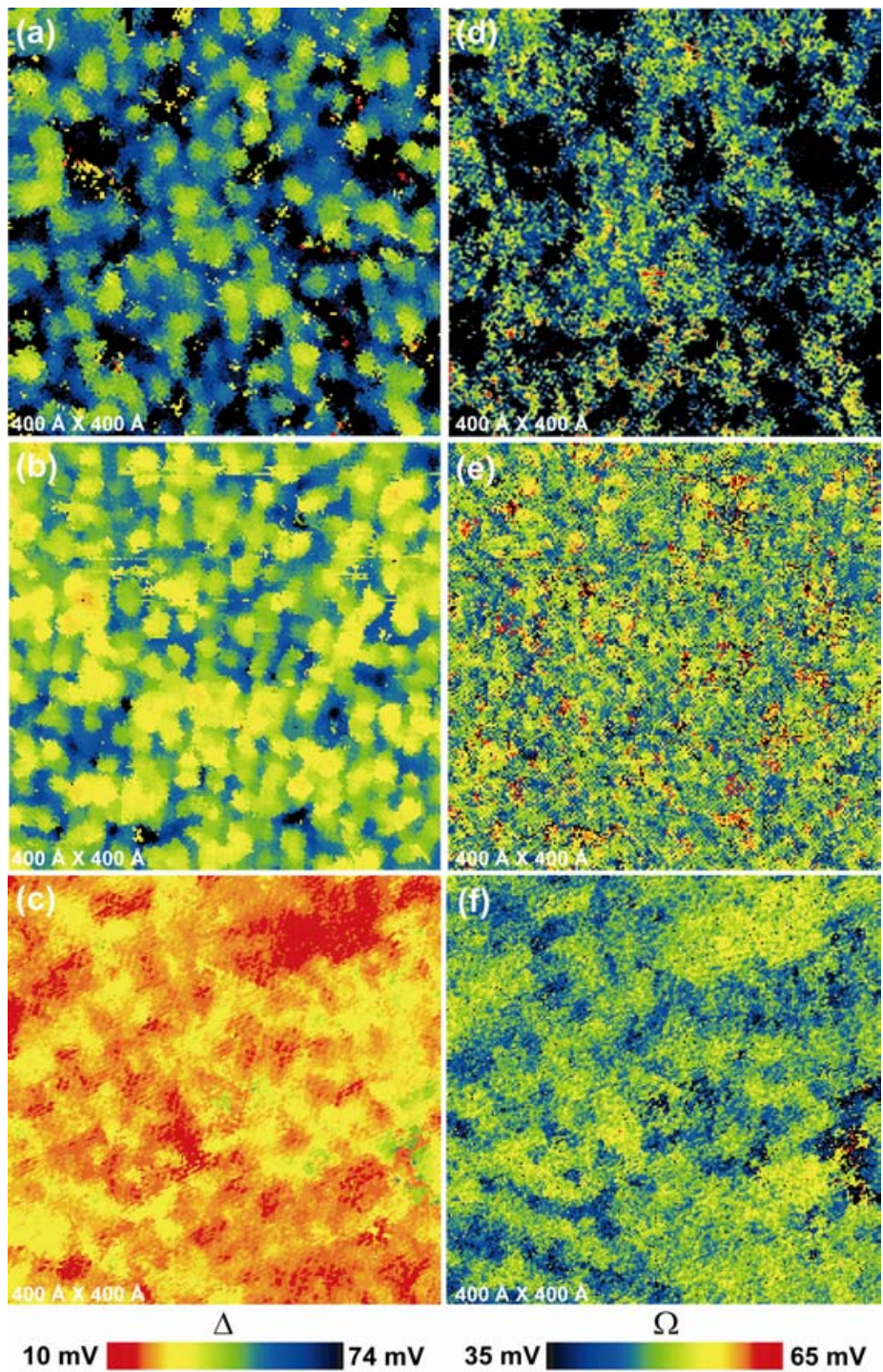

**Figure 3.**

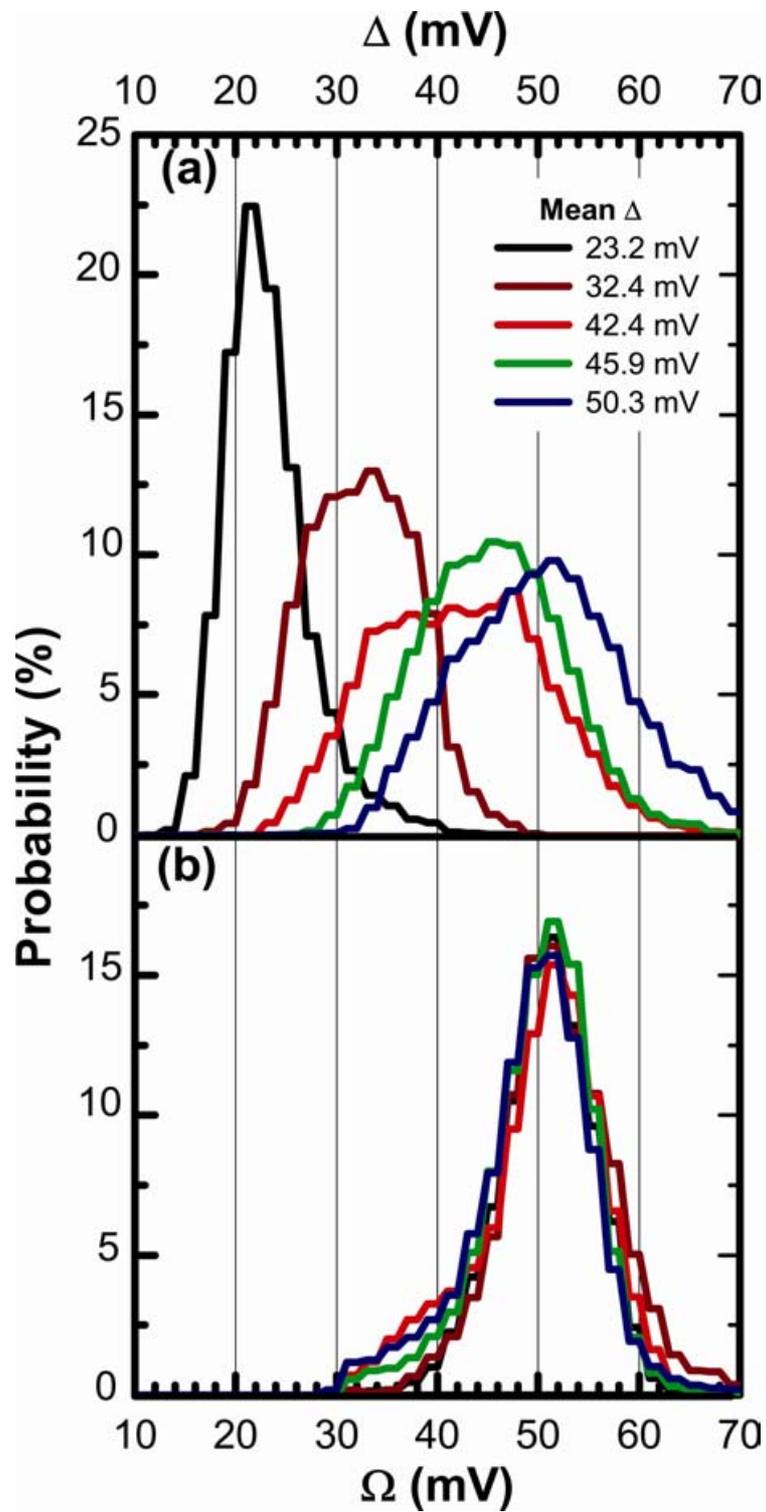

**Figure 4.**

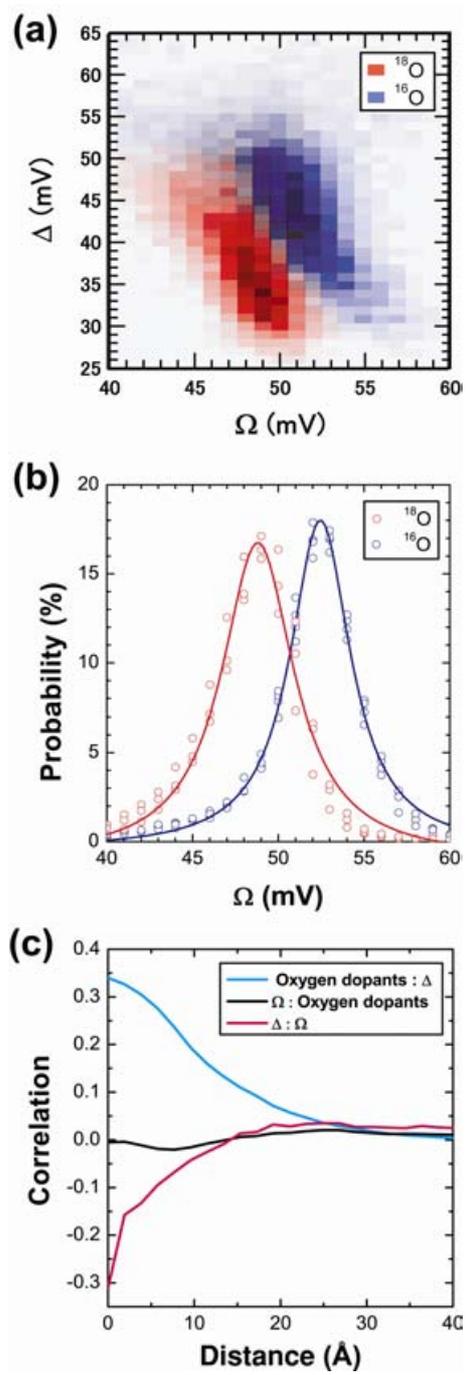

**Figure 5.**

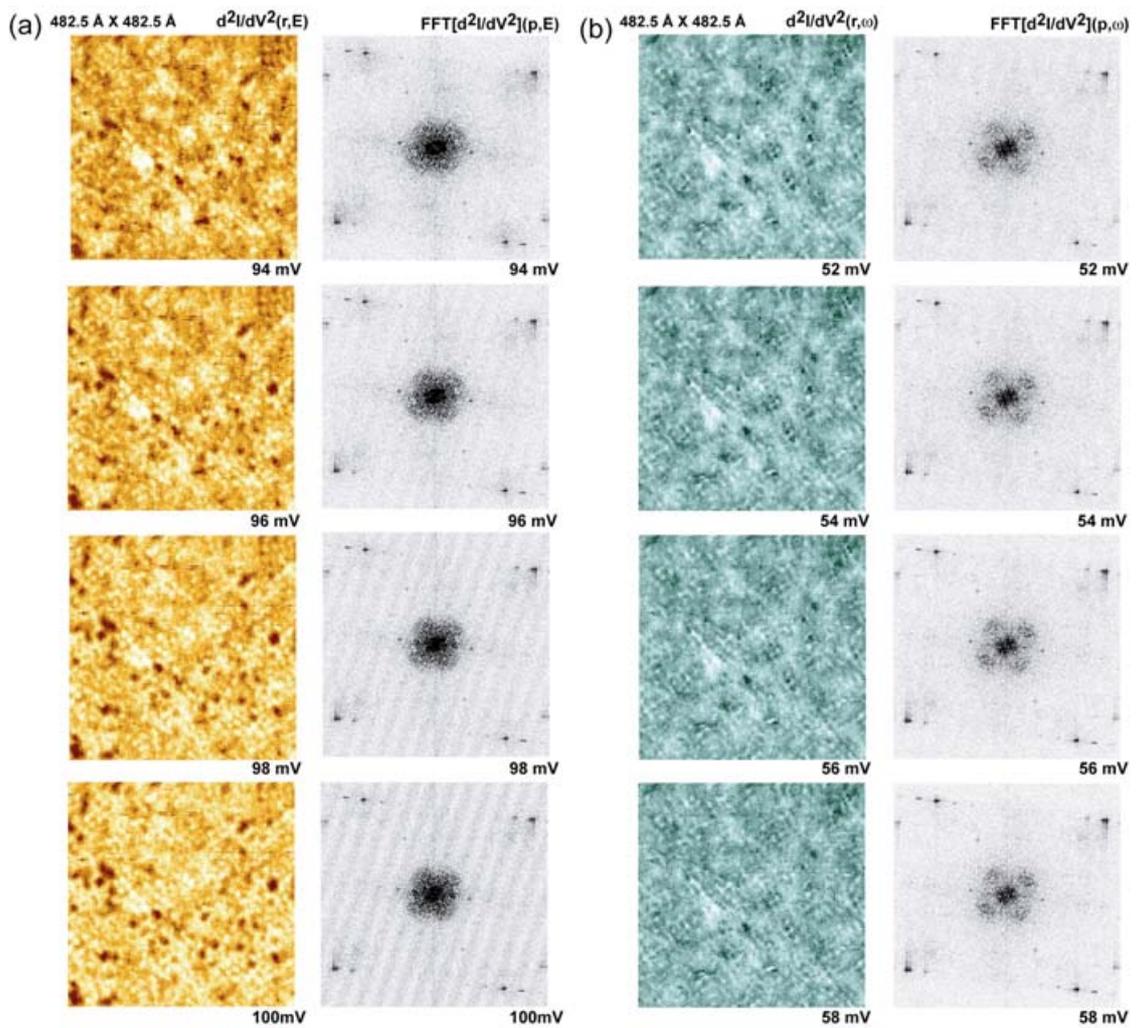

**Supplementary Figure 1.**

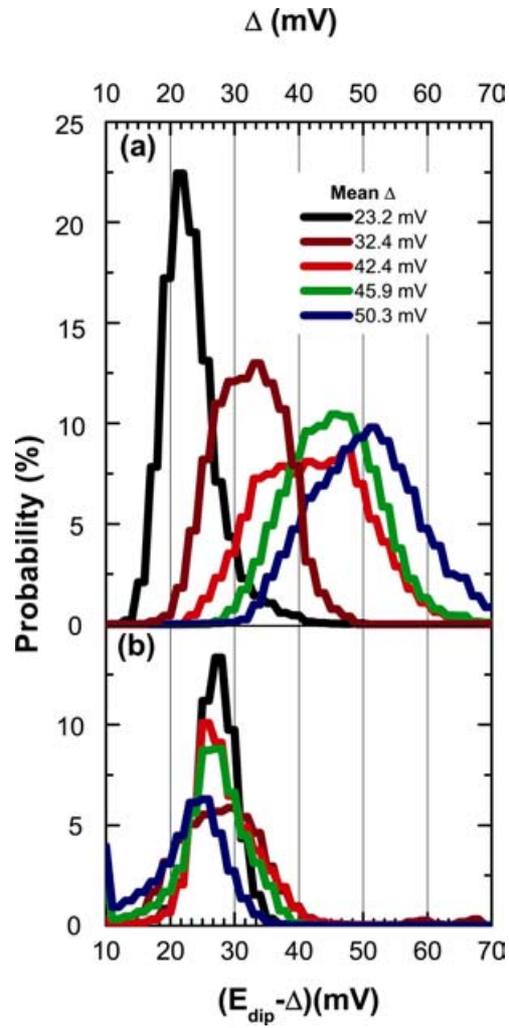

**Supplementary Figure 2.**

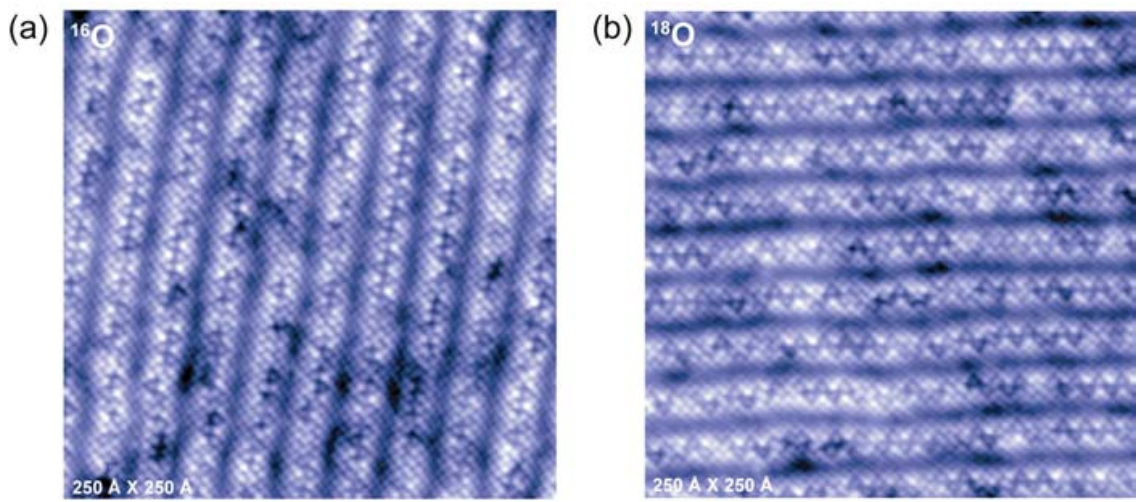

**Supplementary Figure 3**